\documentclass[prl,reprint,amsmath,amssymb,aps,superscriptaddress]{revtex4-2}
\usepackage{xcolor}
\usepackage{graphicx}
\usepackage{hyperref}
\usepackage{physics}
\usepackage{siunitx}
\usepackage{ulem}
\hypersetup{
	colorlinks=true,
	urlcolor=purple,
	citecolor=purple,
	linkcolor=purple,
	breaklinks,
    }

\begin{document}

\title{Probing Entanglement and Symmetries in Random States Using a Superconducting Quantum Processor}

\newcommand{\zju}{School of Physics, ZJU-Hangzhou Global Scientific and Technological Innovation Center, \\and Zhejiang Key Laboratory of Micro-nano Quantum Chips and Quantum Control, Zhejiang University, Hangzhou, China}
\newcommand{\SISSA}{SISSA and INFN, via Bonomea 265, 34136 Trieste, Italy }
\newcommand{\ICTP}{International Centre for Theoretical Physics (ICTP), Strada Costiera 11, 34151 Trieste, Italy}

\author{Jia-Nan Yang} 
\thanks{These authors contributed equally.}
\affiliation{\zju}

\author{Lata Kh Joshi}
\thanks{These authors contributed equally.}
\affiliation{\SISSA}

\author{Filiberto Ares}
\affiliation{\SISSA}

\author{Yihang Han}
\affiliation{\zju}

\author{Pengfei Zhang} 
\email{pfzhang@zju.edu.cn}
\affiliation{\zju}

\author{Pasquale Calabrese}
\email{calabrese@sissa.it}
\affiliation{\SISSA}

\begin{abstract}
Quantum many-body systems display an extraordinary degree of complexity, yet many of their features are universal: they depend not on microscopic details, but on a few fundamental physical aspects such as symmetries. A central challenge is to distill these universal characteristics from model-specific ones. Random quantum states sampled from a uniform distribution, the Haar measure, provide a powerful framework for capturing this typicality. Here, we experimentally study the entanglement and symmetries of random many-body quantum states generated by evolving simple product states under ergodic Floquet models. We find excellent agreement with the predictions from the Haar-random state ensemble. First, we measure the R\'enyi-2 entanglement entropy as a function of the subsystem size, observing the Page curve. Second, we probe the subsystem symmetries using entanglement asymmetry. Finally, we measure the moments of partially transposed reduced density matrices obtained by tracing out part of the system in the generated ensembles, thereby revealing distinct entanglement phases. Our results offer an experimental perspective on the typical entanglement and symmetries of  many-body quantum systems.
\end{abstract}

\maketitle

Originally rooted in random matrix theory and quantum chaos, Page's pioneering work used random states of qubits to model the evaporation of a black hole~\cite{Page1993, Page1993-2}. Page showed that the average entanglement entropy of Haar random quantum states follows a characteristic curve now known as Page curve. Haar random quantum states have since found practical applications in thermalization~\cite{Alessio2016}, quantum information~\cite{Hayden2004, Preskill2007}, complexity theory~\cite{complexity_Haferkamp, complexity_preskill}, quantum cryptography~\cite{Pirandola2020}, and benchmarking quantum processors~\cite{Neill2018, Arute2019}.

While the analytic tractability of Haar-random pure states has resulted in significant advances in the theoretical understanding of universal features of quantum many-body systems, generating and certifying these states in experiments is an exponentially hard task. In practice, it is often sufficient to produce a state $k$-design, namely a finite ensemble of pure states that reproduces the first $k$ statistical moments of the states from a Haar distribution; this observation has motivated extensive research into the efficient preparation of such designs~\cite{cui2025unitary, cotler_kdesign}. Harnessing the intrinsic quantum randomness of projective measurements, recent experiments have demonstrated the preparation of such $k$-designs~\cite{cotler_kdesign, Choi2023}, highlighting their usefulness for benchmarking and characterizing quantum devices. 

Here, we take a practical approach to generate state $k$-designs by explicitly incorporating randomness into the local potentials, an approach motivated by recent realizations of ergodic Floquet models with Wigner-Dyson-like energy level statistics~\cite{Dong2025Phys.Rev.Lett., Joshi2022Phys.Rev.X}. Beginning with a simple product state $\ket{\mathbf{0}}=\ket{0}^{\otimes L}$ in a system of $L$ qubits, we apply a Floquet circuit $V$ at depth $\tau$ to construct states of the type
\begin{equation}
    \ket{\psi}=(V)^\tau \ket{\mathbf{0}}~.
    \label{eq:designStates}
\end{equation}
As shown by our measurements, the resulting random states are, on average, independent of the dynamics that generate them. In particular, this ensemble forms an approximate state $k$-design, as our experimental results closely follow the predictions for Haar-random states for the first few values of $k$.

The Page curve is not the only characteristic feature of Haar-random states. This ensemble is also insightful in the study of entanglement in typical mixed states. References~\cite{PRXQuantum.2.030347, r2_phases} point out distinct entanglement phases when these states are divided into three subsystems. A commonly used measure to distinguish these phases relies on the partial transpose moments of the reduced density matrices of the tri-partitioned state. These moments are accessible in our experiment. 

In combination with entanglement, much recent research has focused on refining our understanding of quantum many-body physics through symmetries~\cite{Laflorencie_2014, Gold_Sela}. In particular, the extent to which a subsystem of an extended quantum system is symmetric or asymmetric with respect to a global symmetry has been extensively studied using \textit{entanglement asymmetry} (EA)~\cite{ares2023entanglement}, both in equilibrium states and in out-of-equilibrium quench dynamics~\cite{ Capizzi_2024, joshi-24, Ares2025Mpemba}. In Haar-random states, EA displays non-trivial behavior~\cite{Ares2024Phys.Rev.D, Russotto2025NonAbelian}. In the thermodynamic limit, one finds a sharp transition from a symmetric to a non-symmetric state at half-system size --- a result consistent with the proposal where one waits half the lifetime of a black hole before making any information recovery~\cite{Preskill2007}. 

Our work introduces low-depth quantum circuits for the preparation of $k$-designs and presents measurements of Page curves for both entanglement entropy and asymmetry, as well as an analysis of the entanglement phases of random states. 
We estimate these quantities using the classical shadow formalism based on randomized measurements~\cite{Huang2020Nat.Phys., Elben2022Nat.Rev.Phys.} in a superconducting quantum processor.

The paper is organized as follows. After introducing the experimental setup and the protocol used to generate random states, we present measurements of the Page curve of the R\'enyi-2 entanglement entropy and for the EA, and compare them with the theoretical predictions for the Haar ensemble. We then study the mixed-state entanglement of the generated states through their partial-transpose moments, experimentally observing the entanglement phase diagram expected for Haar-random states. Finally, we demonstrate how our protocol can also be used to simulate ergodic many-body quantum dynamics by measuring the time evolution of the R\'enyi-2 entanglement entropy and the EA in the implemented Floquet circuit.

\section{Experimental protocol for generating random quantum states}

\begin{figure}[t]
    \centering
    \includegraphics[width=\linewidth]{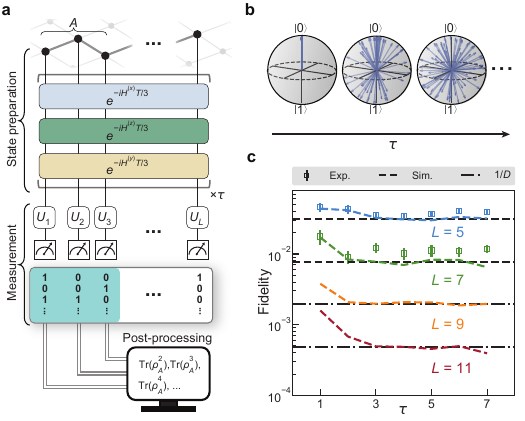}
    \caption{
    {\bf Experimental protocol for generating and characterizing quantum random states.} 
    {\bf a}, Experimental sequence. Floquet driving, Eq.~\eqref{eq:floquet_operator}, is applied for $\tau$ cycles to generate random states. Before performing projective measurements, random single-qubit gates $\{U_{l}\}$, drawn from CUE(2), are applied to each qubit. The measured bitstrings, together with the applied random gates, provide classical shadows of the density matrix. Properties of subsystem $A$ can then be estimated from these classical shadows obtained for various sets of random single qubit gates (see Methods).
    {\bf b}, Visualization of the Floquet dynamics. Quantum states (blue arrows) gradually scramble with the number of Floquet cycles $\tau$, yielding an approximately uniform distribution over the Hilbert space (gray sphere).
    {\bf c}, Average fidelity between the generated random states  as a function of $\tau$. 
    As the number of Floquet cycles increases, the average fidelity converges toward the prediction for the Haar-random ensemble $2^{-L}$ (dash-dotted lines).    Dashed lines: noiseless numerical simulations. Dots and error bars: measured values. 
    }
    \label{fig:fig1_concept}
\end{figure}

To generate the random states, we implemented a periodically kicked spin-$1/2$ system (see Fig.~\ref{fig:fig1_concept}\textbf{a})~\cite{Joshi2022Phys.Rev.X, Dong2025Phys.Rev.Lett.}. From the product state $\ket{\mathbf{0}}$, the system is quenched with a Floquet time-evolution operator $V$, given by
\begin{equation}
    \label{eq:floquet_operator}
    V = e^{-i H^{\left(y\right)}T/3} e^{-i H^{\left(z\right)} T /3} e^{-i H^{\left(x\right)} T /3},
\end{equation}
with $T$ denoting one Floquet cycle.
The constituent Hamiltonians $H^{\left(x, y, z\right)}$ describe the nearest-neighbor interactions in the spin chain with disordered local fields on each site,
\begin{equation}
    H^{\left(x, y, z\right)} = J\sum_{l=1}^{L-1} \left(\sigma^{+}_{l}\sigma^{-}_{l+1} + \sigma^{-}_{l}\sigma^{+}_{l+1}\right) + \sum_{l=1}^{L} h_{l}^{\left(x, y, z\right)} \sigma_{l}^{\left(x, y, z\right)},
\end{equation}
with $J$ denoting the nearest-neighbor coupling strength. The disordered local fields $h_{l}^{\left(x, y, z\right)}$ take random values drawn independently from a uniform distribution over the interval $\left[-J, J\right]$. For a Floquet period $T$ of the order of $1/J$, the dynamics is known to be ergodic~\cite{Joshi2022Phys.Rev.X}.

The experiments were performed on a superconducting quantum processor featuring qubits arranged in a two-dimensional square lattice. Here, the nearest-neighbor coupling strength is set to $J/2\pi=-5$~MHz, and the Floquet period is fixed at $T=90~\unit{\nano\second}$; see Supplementary Material (SM) Section~S1 for details of the experimental setup.
Unitary evolution of an initial product state within such chaotic quantum system induces a rapid scrambling of quantum information. For a sufficiently large number of Floquet cycles $\tau$, the resulting state $V^{\tau} \ket{\mathbf{0}}$ becomes an excellent approximation of a state drawn from the Haar measure, see also~\cite{longtime_rigol, lazarides_14, elben_ndesign}. The uniform state preparation is illustrated in Fig.~\ref{fig:fig1_concept}\textbf{b}, where the many-body Hilbert space is represented by the gray sphere. Initially, all states from the ensemble point in the same direction (up, $\ket{\mathbf{0}}$). With each Floquet cycle under \eqref{eq:floquet_operator}, the states distribute more uniformly over the entire Hilbert space.

For the system sizes studied in this work $(L=5,7,9,$ and $11)$, we find that only a few Floquet cycles are sufficient to generate an approximate state $k$-design that closely approximates the Haar-random behavior for $k=2$, $3$ and $4$ (see SM Section~S2). One verification of this behavior is shown in Fig.~\ref{fig:fig1_concept}\textbf{c}. In the first step, as shown in panel~\textbf{a}, we have created an ensemble of states $\{\rho_r =\ket{\psi_r} \bra{\psi_r}\}$ with $r$ labeling the states. 
In the next step, we estimate the average fidelities between these states from the measurement data, as shown by dots and error bars in panel~\textbf{c}, for system sizes $L=5$ and $7$. The average fidelity $\mathbb E_{\textrm{Haar}}\left[\left|\bra{\phi_r}\ket{\phi_{r'}}\right|^2\right]$ between states $\ket{\phi_r}$, $\ket{\phi_{r'}}$ drawn from the Haar-random ensemble is $1/D$, where $D$ denotes the corresponding Hilbert space dimension, here $D=2^L$. In the noiseless numerical simulations of the protocol, corresponding to the colored dashed curves, we find that our ensembles for different $L$ tend to $1/D$ (dot-dashed lines) after $\tau=3$ cycles. However, estimating such small fidelities experimentally is a challenging task, as it requires a very large number of measurement bases. For this reason, we rely on numerical simulations for larger system sizes $L=9, 11$. Based on these results, in the following sections we choose  $\tau=7$ as the optimal number of Floquet cycles for generating an ensemble of states that closely approximates Haar randomness.

\section{R\'enyi entanglement entropy} 

Entanglement entropy is one of the most fundamental quantities for probing quantum entanglement in pure states. Consider a chain divided into two contiguous subsystems $A$ and $B$. The R\'enyi-2 entropy of subsystem $A$ is defined as
\begin{equation}
    S_{A}^{\left(2\right)} = -\log\left(\text{Tr}\left[\rho_{A}^2\right]\right),
\end{equation}
with $\rho_{A}$ being the reduced density matrix of $A$.
Experimentally, the R\'enyi entropy can be measured via different techniques including, for example, two-copy state preparation~\cite{daley2012measuring, islam2015measuring}, quantum state tomography (QST)~\cite{steffen2006measurement, de2013complete}, and randomized measurements (RMs)~\cite{Brydges2019Science, Satzinger2021Science}. 
For moderate sized subsystems, RMs are significantly more efficient than QST or other approaches because they require fewer measurements and only a single copy of the quantum state. The circuit used to implement classical shadows --- a specific RM protocol --- is shown in Fig.~\ref{fig:fig1_concept}\textbf{a}, where random single-qubit gates $\{U_l\}$, drawn from the 2-dimensional circular unitary ensemble (CUE), are applied prior to the projective measurement onto the computational basis of each qubit. For each set of random single-qubit gates, we collect $K$ measurement shots and use all outcomes to construct classical shadows of the unknown density matrix. These are sufficient to estimate the R\'enyi-2 entropy (see Methods and SM Section~S3 for details).

\begin{figure}[t]
    \centering
    \includegraphics[width=\linewidth]{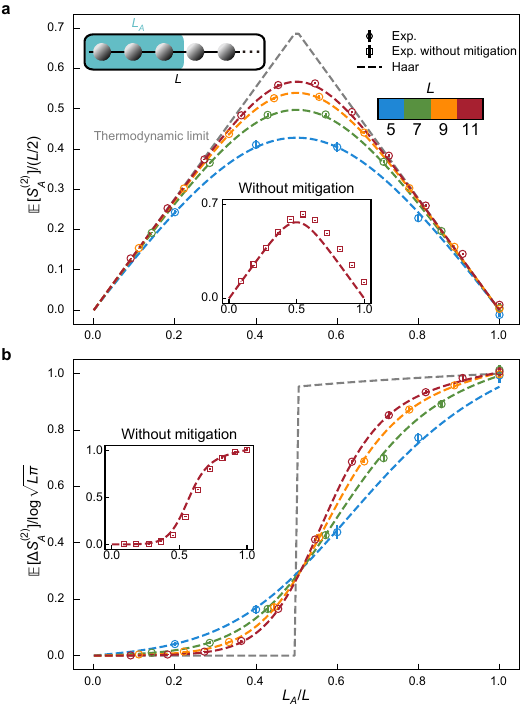}
    \caption{
    {\bf Experimental Page curves of entanglement entropy and entanglement asymmetry.}
    Measurements of the average R\'enyi-2 entanglement entropy $\mathbb{E}[S_A^{(2)}]$ ({\bf a}) and entanglement asymmetry $\mathbb{E}[\Delta S_A^{(2)}]$ ({\bf b}) as functions of the subsystem size $L_A$ are shown for different total system sizes $L$. The dashed curves are the theoretical prediction for Haar-random states for the corresponding $L$, while the gray dashed curves indicate the thermodynamic limit $L\to\infty$. In the insets, the symbols are the measured values for system size $L=11$. We observe deviations from the theoretical expectations (dashed curves) due to decoherence during the Floquet evolution that generates the random states, impacting the unitarity of the evolution. These errors can be mitigated as explained in the main text and SM. In the main panels, the symbols are the error-corrected results, which perfectly follow prediction for the Haar-random ensemble. 
    Error bars denote the standard error of the mean (SEM) over different states (see SM Section~{S3E} for the error bars due to finite number of measurement bases). We average over $15$ random states for $L = 5,7$ and $10$ random states for $L= 9,11$. 
    }
    \label{fig:fig2_RE_EA}
\end{figure}
In Fig.~\ref{fig:fig2_RE_EA}\textbf{a}, we show the measurements of the averaged R\'enyi-2 entropy $\mathbb E[S_A^{(2)}]$ for the states generated in the experiment. Throughout the text, $\mathbb{E}[\cdot]$ denotes the averaged value of the observable $\cdot$ over the ensemble of random states. We plot the entropy as a function of the subsystem size $L_A$, for various total system sizes $L$. The dashed lines correspond to the averaged R\'enyi-2 entanglement entropy in the Haar-random ensemble (that is, the Page curve) for the corresponding system sizes; see SM Section~S4 for analytic formulas for observables in the Haar ensemble. The readout errors, occurring during the projective measurements, are corrected using a response matrix-based method~\cite{Bravyi2021Phys.Rev.A, PhysRevLett.118.210504} (see SM Section~{S3C}). The measurements of the entropy after this correction are shown in the inset for system size $L=11$. We see that the measured results show a systematic deviation from the theoretical predictions. This deviation grows with subsystem size and is particularly evident for the full system, where the entropy fails to vanish, contrary to what is expected for a pure state. This behavior arises from experimental decoherence during the Floquet evolution, which effectively renders the dynamics open and leads to the generation of a mixed state. To address this issue, we implement a comprehensive error mitigation protocol. The dynamics governed by the Floquet operator $V$ in Eq.~\eqref{eq:floquet_operator} is that of a CUE matrix. The average error during such Haar random evolutions behaves like a depolarization channel~\cite{Mele2024introductiontohaar}, and thus the prepared random state at any cycle $\tau$ is $\rho (\tau)\rightarrow (1-\epsilon)\rho(\tau)+\epsilon (\tau)  \,\mathbb{I}/D$. The decoherence rate $\epsilon$  can be estimated from the experimental data itself. To have an unbiased result, we divide our ensemble into two halves. Half of the states are used to get the value $\epsilon$, and the other half are used to obtain the average error-corrected results of entropy and EA (see more details in the SM Section~{S3D}).

After applying this error mitigation procedure to account for decoherence, the corrected results correctly reproduce the Haar-random predictions, as seen in Fig.~\ref{fig:fig2_RE_EA}\textbf{a}. The states we generate are highly entangled, and their averaged entanglement entropy follows a volume law, see also \cite{CHOWDHURY2025117112}. In particular, as expected for Haar-random states, it grows approximately at the maximal rate, namely, $L_A \log 2$ for $L_A < L/2$ when $L$ is large enough.  The entropy then decreases symmetrically beyond the half partition, as is the characteristic of pure states.
We emphasize that  observing the Page curve in the experimentally created state ensembles implies that these states are, on average, independent of the specific details used to prepare them. 

\section{Entanglement asymmetry}  

A key advantage of the classical-shadow protocol is that the same measurement dataset used to estimate the Rényi-2 entanglement entropy also provides access to a broad class of observables. We exploit this capability to probe the symmetries of the random-state ensemble realized in our experiment via the EA, which quantifies the symmetry breaking within a subsystem $A$~\cite{ares2023entanglement, joshi-24}.

Although Haar-random states lack any global symmetry, this does not necessarily imply that symmetries are broken at the level of a subsystem. Without loss of generality, we consider the $U(1)$ symmetry generated by the charge $Q = \sum_{l=1}^L \ket{1}_l \bra{1}_l$, which counts the total number of qubits in the state $\ket{1}$. This charge decomposes into the contribution from subsystem $A$ and its complement, $Q=Q_A+Q_B$. The reduced density matrix $\rho_A$ preserves the symmetry if $[\rho_A, Q_A] = 0$; otherwise, the symmetry is broken. The (R\'enyi-2) EA~\cite{ares2023entanglement} is defined as
\begin{equation}
    \Delta S_{A}^{(2)} = S_{A,Q}^{(2)}-S_{A}^{(2)},
    \label{Eq:EA}
\end{equation}
where $S_{A,Q}^{(2)}$ is the R\'enyi-2 entropy of the symmetrized state $\rho_{A, Q}$ obtained from $\rho_A$ via $\rho_{A,Q} = \sum_{q \in \mathbb{Z}} \Pi_{q} \rho_{A} \Pi_{q}$, with $\Pi_{q}$ projecting onto the eigenspace of $Q_A$ with charge $q \in \mathbb{Z}$. By construction, $[\rho_{A, Q}, Q_A]=0$, and, when the symmetry is preserved in $A$, $\rho_{A, Q}=\rho_A$. Therefore, the EA vanishes if and only if $\rho_A$ is symmetric.

Figure~\ref{fig:fig2_RE_EA}\textbf{b} shows the average EA of the ensemble of states generated in the experiment when varying $L_A$ for several sizes $L$ of the total system. In the main plot, symbols are the experimental results after applying the same error mitigation protocol as for the entanglement entropy. The results  before applying the error mitigation for $L=11$ are shown in the inset. The dashed lines are the theoretical average EA for the Haar-random state ensemble~\cite{Ares2024Phys.Rev.D}. Notice that, remarkably, when comparing with the entanglement entropy in Fig.~\ref{fig:fig2_RE_EA}\textbf{a}, the effects of decoherences are less prominent in the EA. This is likely because the EA is a difference between entropies and, as a result, some systematic errors in the individual entropies can partially cancel out when taking the difference. 

The EA grows monotonically with $L_A$, signaling that the $U(1)$ symmetry generated by $Q_A$ is increasingly broken with the subsystem size. For subsystems of size $L_A < L/2$, the EA decreases with $L$ and eventually vanishes in the thermodynamic limit $L\to \infty$ (gray dashed curve); therefore, these subsystems respect the $U(1)$ symmetry in the large-$L$ limit. By contrast, for $L_A > L/2$, the EA rises with $L$ and approaches $1/2\log(\pi L_A)$ as $L\to\infty$. Near $L_A=L/2$, the EA varies rapidly and develops a jump discontinuity in the thermodynamic limit, corresponding to a sudden transition of $\rho_A$ from a symmetric to a non-symmetric state. This transition is a characteristic feature of Haar-random states for any symmetry group, including non-Abelian ones~\cite{Russotto2025NonAbelian}. 
Within the Page model of an evaporating black hole, this result implies that the emitted radiation is symmetric before the Page time ($L_A = L/2$) and asymmetric thereafter.

\section{Entanglement phase diagram} 
\begin{figure*}[ht!]
    \centering
    \includegraphics[width=\linewidth]{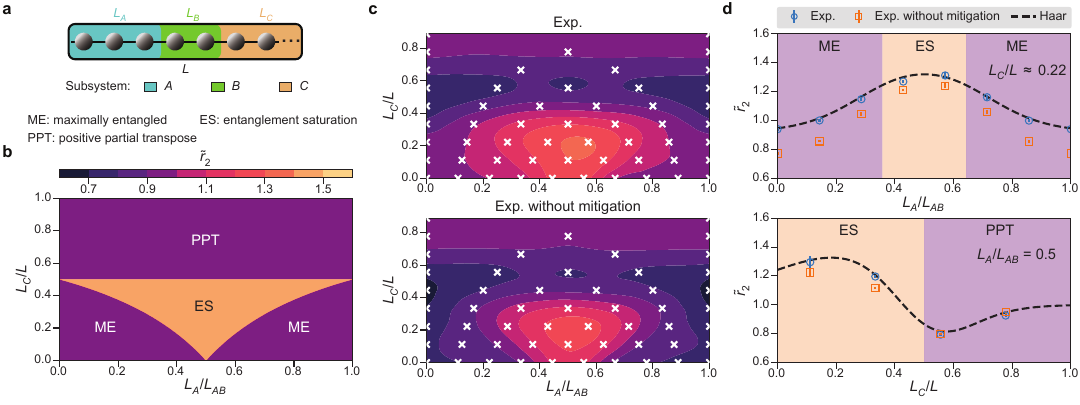}
    \caption{
    \textbf{Entanglement phase diagram from partial transpose moments.} 
    \textbf{a}, Schematic illustration of partitioning the whole qubit chain into three subsystems: $A$ (blue), $B$ (green), and $C$ (yellow). Subsystem $C$ is taken as the environment, and the partial transpose with respect to $B$ is computed for the reduced density matrix of $AB$. 
    \textbf{b}, Theoretical phase diagram expected for the Haar-random state ensemble using the ratio $\tilde{r}_2$, defined in Eq.~\eqref{eq:r2}, in the thermodynamic limit $L\to\infty$. The asymptotic values are $\tilde{r}_2 = 1$ for the ME and PPT phases, and $\tilde{r}_2 = 3/2$ for the ES phase. 
    \textbf{c}, Experimental phase diagram of the ratio $\tilde{r}_2$ for a finite system size $L = 9$. The top and bottom panels show the error-mitigated and raw results, respectively. White crosses indicate measured data points, used to draw the colored contour plot. 
    \textbf{d}, Horizontal (top) and vertical (bottom) slices of the phase diagram. The horizontal slice corresponds to $L_C / L \approx 0.22$, and the vertical slice is taken at $L_A / L_{AB} = 0.5$. Circles and squares represent error-mitigated and raw data, respectively. Dashed lines indicate theoretical predictions for the Haar-random ensemble of $L = 9$. Error bars are obtained via jackknife resampling (see SM Section~{S3E}).
    }
    \label{fig:fig3_r2}
\end{figure*}

So far, we have studied entanglement  and symmetries in pure states. Our measurements also enable the study of properties of \textit{mixed} random states. To investigate them, we partition the system into three subsystems: $A$, $B$, and $C$ (Fig.~\ref{fig:fig3_r2}\textbf{a}). By treating the subsystem $C$ as an environment, the remaining state defined on the subsystems $A$ and $B$ is described by the reduced density matrix $\rho_{AB} = \operatorname{Tr}_C[\rho]$. When the full system is in a Haar-random state, this tripartite configuration displays different entanglement phases for different subsystem sizes.
 Two established metrics for characterizing such transitions are the logarithmic negativity~\cite{PRXQuantum.2.030347} and the ratio of the partially transposed moments of $\rho_{AB}$~\cite{r2_phases}. Here, we adopt the latter approach because it is defined in terms of only a few low-order moments of $\rho_{AB}$, making it highly amenable to efficient experimental measurements.

The partial transpose of the reduced density matrix $\rho_{AB}$ with respect to subsystem $B$ is defined as $\rho^{\Gamma} = \left(\mathbb{I}_A \otimes T_B\right) \rho_{AB}$, where $\mathbb{I}_A$ is the identity operator on subsystem $A$ and $T_B$ is the transposition operator on subsystem $B$. The ratio $\tilde{r}_2$,
\begin{equation}\label{eq:r2}
    \tilde{r}_2 = \frac{\mathbb{E}[p_2] \mathbb{E}[p_3]}{\mathbb{E}[p_4]},
\end{equation}
is the lowest order ratio that sufficiently distills different entanglement phases. Here, $p_n = \operatorname{Tr}\left[ \left(\rho^{\Gamma}\right)^n \right]$ denotes the $n$th-order moment of the partially transposed state $\rho^{\Gamma}$. Notice that we have retained the notation $\tilde{r}_2$, as in Ref.~\cite{r2_phases}, to distinguish it from the ratio $r_2=p_2p_3/p_4$ defined for a single quantum state.

For Haar-random states, in the thermodynamic limit, three distinct entanglement phases have been identified: the maximally entangled (ME) phase where $\tilde r_2=1$, the entanglement saturation (ES) phase where $\tilde r_2=3/2$, and the positive partial transpose (PPT) phase where $\tilde r_2=1$, as illustrated in Fig.~\ref{fig:fig3_r2}\textbf{b}. Although the ME and PPT phases share the same value $\tilde{r}_2 = 1$, they can be further distinguished using the $p_3$-negativity~\cite{r2_phases} (see SM Section~{S5B}).
The phase diagram exhibits several particularly interesting aspects. First, in the absence of an environment subsystem ($L_C = 0$), the system remains in the ME phase, showing $\tilde r_2$=1 for all bipartitions of $AB$. This is the bipartite case studied previously with the R\'enyi entropy, which is described by the Page curve.  
Second, when the environment $C$ is smaller than $AB$ ($L_C < L_{AB}$, where $L_{AB} = L_A + L_B$), $\rho_{AB}$ has a non-positive partial transpose, and entanglement shows different features depending on the partition between $A$ and $B$. For $L_A > L_{BC}$ or $L_B > L_{AC}$ (with $L_{BC} = L_B + L_C$ and $L_{AC} = L_A + L_C$), subsystems $A$ and $B$ are maximally entangled --- hence the name ME phase --- and the smaller of $A$ and $B$ is not entangled with $C$. In contrast, near $L_A \simeq L_B$, $A$ and $B$ are no longer maximally entangled; instead, there is mutual entanglement between $A$, $B$ and $C$, and the system enters the ES phase. Consequently, for a fixed environment size $L_C<L_{AB}$, varying the subsystem size $L_A$ drives a transition between the ME and ES phases. Finally, for $L_C>L_{AB}$, $\rho_{AB}$ has a positive partial transpose. In particular, fixing the ratio $L_A/L_{AB} = 0.50$ and increasing $L_C$, $\rho_{AB}$ undergoes a transition from the ES phase to the PPT phase at $L_C=L/2$.

Experimental results for a total system size of $L=9$, both with and without error mitigation, are shown in Fig.~\ref{fig:fig3_r2}\textbf{c}-\textbf{d}. The estimation of $p_3$ and $p_4$ moments is computationally demanding,  therefore we have used the method of batch shadows~\cite{PRXQuantum.4.010318}, which reduces the post-processing time significantly (see Methods). The shaded region in Fig.~\ref{fig:fig3_r2}\textbf{c} was obtained by a cubic interpolation of the experimental data (white crosses) to have a smooth contour plot. Unlike the thermodynamic limit, where $\tilde r_2$ takes constant values within each phase and exhibits sharp jumps between them, in finite-size systems the phase boundaries are smeared and $\tilde{r}_2$ varies smoothly, making phases identifiable through the dominant behavior.
The transition between ME and ES phases is shown in the top panel of Fig.~\ref{fig:fig3_r2}\textbf{d} by fixing $L_C/L \approx 0.22$, and the transition between ES and PPT regions for $L_A/L_{AB}=0.50$ in the lower panel. Dashed lines correspond to the theoretical prediction for the Haar-random ensemble at finite $L$.

In SM Section~{S5A}, we present the phase diagram for system size $L=11$. In particular, in this case, the estimation of the moments requires a large measurement budget, which we statistically overcome by considering averages over different choices of subsystems.
Remarkably, the experimentally obtained phase diagram for both $L=9$ and $L=11$ closely matches the Haar-random prediction in Fig.~\ref{fig:fig3_r2}\textbf{b}. Therefore, our protocol enables to map out experimentally the full phase diagram for approximate $k$-design states.

\section{Out of equilibrium dynamics}

\begin{figure}[t]
    \centering
    \includegraphics[width=\linewidth]{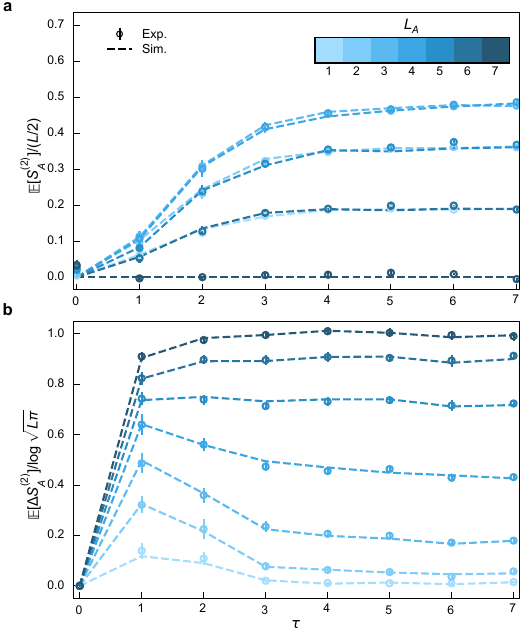}
    \caption{
    \textbf{Dynamics of R\'enyi-2 entanglement entropy (a) and entanglement asymmetry (b) in the Floquet circuit.}
    Each experimental data point is the error-corrected average over 10 random realizations of the circuit, for various sizes $L_A$ of subsystem $A$ and total system size $L=7$.  The error bars represent the SEM. Dashed curves were obtained from the noiseless numerical simulation of the Floquet circuit.
    }
    \label{fig:fig4_EA_VS_tau}
\end{figure}

Until now, we have examined the approximated random states obtained after $\tau=7$ Floquet cycles. The same circuit can also be used to experimentally probe the properties of ergodic dynamics. Figure~\ref{fig:fig4_EA_VS_tau} shows the time evolution of the average Rényi-2 entanglement entropy (panel \textbf{a}) and EA (panel \textbf{b}) as functions of $\tau$ for various subsystem sizes $L_A$ in a system of $L=7$ qubits.

As expected for ergodic dynamics starting from a product state, the entanglement entropy grows at early times and saturates near the value expected from the Haar-random ensemble~\cite{Kim2013, Chan2018}. The EA is initially zero since the initial state $\ket{\mathbf{0}}$ is an eigenstate of the charge $Q$ and, therefore, is symmetric. 
The Floquet circuit~\eqref{eq:designStates} does not commute with $Q$, and the total state $\ket{\psi}$ does not respect the symmetry.
At late times, the EA saturates to the prediction from the Haar-random ensemble,  exhibiting two distinct regimes during time evolution (cf. Fig.~\ref{fig:fig2_RE_EA}\textbf{b}). For $L_A>L/2$, the EA rapidly equilibrates to the late time prediction, whereas for $L_A<L/2$ it peaks at $\tau=1$ before decaying to the stationary value, with the symmetry being restored in the large-$L$ limit. 
This behavior is qualitatively similar to that observed in both local and non-local circuits composed of two-qubit Haar-random gates~\cite{Ares2025RUC}, where the long-time stationary regime is also characterized by the Haar-random ensemble. In those models, the EA relaxation time is independent of the system size for 
$L_A>L/2$, but for $L_A<L/2$ it is governed, similarly to the entanglement entropy, by the scrambling time of the dynamics in the large-$L$ limit. The results in Fig.~\ref{fig:fig4_EA_VS_tau} are consistent with these features. Still, to probe them, it is necessary to consider larger (sub)systems and apply methods that enhance the efficiency of classical shadows (whose computational complexity grows exponentially in the (sub)system size) at large scales~\cite{BV_Piroli, votto2025learningmixedquantumstates}.

\section{Discussion and outlook}

In summary, we employ a Floquet circuit to generate random quantum states on a superconducting quantum processor and investigate their entanglement, through the R\'enyi-2 entanglement entropy and partial-transpose moments, and symmetry properties, 
via the entanglement asymmetry. Although the circuit is shallow, with each Floquet cycle consisting only of nearest-neighbor two-body interactions and local random driven potentials, the generated states clearly exhibit the characteristics of Haar-random states. We experimentally measured the Page curves of the entanglement entropy and asymmetry. We also demonstrated the entanglement phases expected for this ensemble. These results highlight the versatility of our superconducting quantum processor for studying complex many-body quantum systems through the use
of efficient, randomized-measurement-based protocols.
We further experimentally measured the time evolution of the entanglement entropy and asymmetry in the Floquet circuit, observing a similar qualitative behavior to that exhibited by Haar-random circuits.

Our work lays the groundwork for probing  the universal features of generic many-body quantum dynamics in quantum simulators, providing a small-scale example that can be scaled up through combined measurement protocols for larger systems~\cite{BV_Piroli, votto2025learningmixedquantumstates}. Furthermore, this represents a fundamental step toward the experimental investigation of {\it generic} quantum many-body ergodic dynamics, going beyond the study of specific microscopic models and enabling access to universal features of chaotic quantum systems.

While Haar-random states are completely structureless and do not respect any symmetry constraints, recent studies have shown that the presence of additional global symmetries can have a significant impact on the Page curve, entanglement phases, and EA~\cite{bianchi_19, PhysRevD.106.046015, Lau:2022symmetryPageCurve, PhysRevA.106.052428, russotto_25}. It would be interesting to extend the Floquet protocol implemented here to generate ensembles of random states endowed with specific global symmetries. Such states could be created by simulating periodically driving Hamiltonians that respect the desired symmetry. This extension would open the door to experimental investigations of chaotic many-body dynamics in the presence of conservation laws.

\section{Methods}

\noindent{\bf Classical shadows.} Once the quantum state of interest is prepared, a random rotation $U^{(m)}=U_1^{(m)}\otimes U_2^{(m)}\otimes\cdots \otimes U_L^{(m)}$ is applied, where the single-qubit random gates are sampled uniformly and independently from the circular unitary ensemble CUE(2). This is followed by projective measurements on each qubit. This process is repeated $K$  times (also referred to as $K$ shots) for the same set of the random rotations. The measurements are then repeated for $m=1, 2, \dots, M$  different random rotations.  The classical shadow of the state on the subsystem $A$, for each set of random rotations and a single shot, is given by
\begin{equation}
    \hat{\rho}_{A}^{\left(m, k\right)} =\bigotimes_{l=1}^{L_{A}}\left( 3\left( U_{l}^{\left(m\right)} \right)^\dagger \ket{s_{l}^{\left(m, k\right)}} \bra{s_{l}^{\left(m, k\right)}} U_{l}^{\left(m\right)} - \mathbb{I}\right),
\end{equation}
where $s_{l}^{\left(m, k\right)} \in \{0, 1\}$ is the measurement outcome for the $l$-th qubit  from the $k$-th shot in the $m$-th instance of the random rotation. This operator makes an unbiased estimation of the reduced density matrix $\rho_A$, such that the average over random rotations and shots is equal to $\rho_A$, i.e., $\mathbb E [\rho_{A}^{\left(m, k\right)} ]=\rho_A$.  Using these classical shadows, we can accurately estimate the $n$th-order moments of the density matrix using a U-estimator~\cite{Huang2020Nat.Phys.}, correlating the measurement results from different random rotations, as follows
\begin{equation}
\text{Tr}[\rho_A^n]=\frac{\left(M-n\right)!}{M!} \sum_{m_1\neq m_2\dots \neq m_n}\text{Tr}[\hat\rho_A^{(m_1)}\hat\rho_A^{(m_2)}\dots \hat\rho_A^{(m_n)}]~,
\label{Eq:shadow_RE}
\end{equation}
where the operator $\hat\rho^{(m)}=\mathbb E_{K}[\hat\rho^{(m, k)}]$ is the classical shadow corresponding to the $m$-th random rotation constructed from averaging $K$ shots. 
From Eq.~\eqref{Eq:shadow_RE}, the R\'enyi entropy is estimated. The above classical shadows and estimators are directly generalized to obtain the moments of the symmetrized density matrix $\rho_{A, Q}$ used in the entanglement asymmetry, and those of the partial transpose $\rho^\Gamma$ used in the entanglement phase estimation. 

\noindent{\bf Batch shadows.} 
The above estimator quickly becomes computationally impractical for $n\geq 3$ as the system size increases. This is because it involves summing over all distinct combinations of the form $\hat{\rho}_A^{(m_1)} \dots \hat{\rho}_A^{(m_n)}$, where each $m_i$ runs from $1$ to $M$, causing the runtime to scale as the $n$-th power of the number of measurements $M$. As a result, the post-processing becomes unfeasible. To reduce the post-processing runtime when estimating higher moments such as $p_3$ and $p_4$ using classical shadows, we employ the \textit{batch shadows} technique introduced in Ref.~\cite{PRXQuantum.4.010318}. The main idea is to average a subset (batch) of classical shadows before using them to estimate physical quantities. Specifically, we define the $b$-th batch shadow as
\begin{equation}
    \tilde{\rho}_A^{(b)}=\frac{1}{M^\prime}\sum_{m=(b-1)M^\prime+1}^{bM^\prime} \hat{\rho}_A^{(m)},
\end{equation}
where $M'$ is the number of classical shadows included in each batch. The total number of batches is then $B = \left\lfloor M/M' \right\rfloor$ with $M$ denoting the total number of classical shadows.
Once the batch shadows are constructed, estimates of physical observables can be obtained by replacing the original classical shadows with the batch shadows in the corresponding estimators. For example, the second R\'enyi entropy of subsystem $A$ can be estimated as
\begin{equation}
    S_{A}^{(2)} = -\log \left(\frac{1}{B\left(B-1\right)} \sum_{b \neq   b^\prime} \operatorname{Tr} \left[ \tilde{\rho}_{A}^{\left(b\right)} \tilde{\rho}_{A}^{(b^\prime)}\right]\right).
\end{equation}
For all estimations in this work, we use a total of $B=20$ batches. 

\vspace{10pt}

\noindent {\bf Acknowledgments} 
We thank the Superconducting Quantum Computing Group at Zhejiang University for supporting the device and the experimental platform on which the experiment was carried out.
F.A. and P.C. thank Sara Murciano, Lorenzo Piroli, and Angelo Russotto for collaborations on closely related topics, in particular those reported in Refs.~\cite{ares2023entanglement,Ares2025RUC,Russotto2025NonAbelian}.
P.Z. acknowledges support from 
the National Natural Science Foundation of China (Grant no.~12404574) and the Zhejiang Leading Goose (Lingyan) Project (Grant no.~2026C02A2004).
L.K.J. acknowledges support from the European Union’s Horizon Europe program under the Marie Sklodowska Curie Action Project ETHOQS (Grant no. 101151139). P.C. and F.A. acknowledge support from the European Research Council under the Advanced Grant no. 101199196 (MOSE).

\noindent {\bf Author contributions} L.K.J. proposed the ideas and conducted the numerical analysis with theory support from F.A. under the supervision of P.C.; J.-N.Y. carried out the experiments and analysed the experimental data under the supervision of P.Z.; 
L.K.J, F.A., P.Z., and J.-N.Y. co-wrote the manuscript.
All authors contributed to the discussions of the results.

\bibliography{references}

\end{document}


\title{Supplementary Material for ``Probing Entanglement and Symmetries in Random States Using a Superconducting Quantum Processor''} 

\newcommand{\zju}{School of Physics, ZJU-Hangzhou Global Scientific and Technological Innovation Center, \\and Zhejiang Key Laboratory of Micro-nano Quantum Chips and Quantum Control, Zhejiang University, Hangzhou, China}
\newcommand{\SISSA}{SISSA and INFN, via Bonomea 265, 34136 Trieste, Italy }
\newcommand{\ICTP}{International Centre for Theoretical Physics (ICTP), Strada Costiera 11, 34151 Trieste, Italy}

\author{Jia-Nan Yang} 
\thanks{These authors contributed equally.}
\affiliation{\zju}

\author{Lata Kh Joshi}
\thanks{These authors contributed equally.}
\affiliation{\SISSA}

\author{Filiberto Ares}
\affiliation{\SISSA}

\author{Yihang Han}
\affiliation{\zju}

\author{Pengfei Zhang} 
\email{pfzhang@zju.edu.cn}
\affiliation{\zju}

\author{Pasquale Calabrese}
\email{calabrese@sissa.it}
\affiliation{\SISSA}

\maketitle

\setcounter{equation}{0}
\setcounter{figure}{0}
\setcounter{table}{0}
\setcounter{section}{0}
\setcounter{page}{1}
\renewcommand{\theequation}{S\arabic{equation}}
\renewcommand{\thefigure}{S\arabic{figure}}
\renewcommand{\thetable}{S\arabic{table}}
\renewcommand{\thesection}{S\arabic{section}}

\tableofcontents

\section{Experimental setup}

In the experiment, we employ a chain of superconducting qubits with system sizes up to $L=11$. Each qubit is individually addressable, enabling the implementation of arbitrary local drivings: flux control for $Z$-type driving and microwave pulses for $XY$-type driving. Nearest-neighbor qubits are coupled via a tunable coupler, which enables precise adjustment of the effective qubit–qubit coupling by tuning the coupler frequency. The coupling strength can be continuously varied within the range $[-20,0]$ MHz. This setup facilitates the exploration of quantum many-body phenomena in a programmable way. The key performance metrics of each qubit, including qubit lifetime $T_1$, spin-echo time $T_2^\mathrm{SE}$, single-qubit gate error $e_\mathrm{sq}$, and readout errors $e_{0\to1}$ and $e_{1\to0}$ are detailed in Table~\ref{tab:T1T2}.

\begin{table*}[!htbp]
    \setlength{\tabcolsep}{12pt}
    \renewcommand{\arraystretch}{1.5}
    \centering
    \caption{
    {\bf Qubit performance.} The qubit lifetime $T_{1}$ and spin-echo coherence time $T_{2}^\mathrm{SE}$ are measured at the interaction frequency $\omega_\mathrm{I}/2\pi = 4.415~\unit{\giga\hertz}$. The single-qubit gate error $e_{\mathrm{sq}}$ of each qubit is measured at its idle frequency via randomized benchmarking. The readout error $e_{j \to k}$ is defined as the probability of measuring the state $\ket{k}$ when the qubit is prepared in the state $\ket{j}$.
    }
    \begin{tabular}{c|c|c|c|c|c}
        \hline \hline
        \multirow{1}{*}{Qubit}
        &  \multirow{1}{*}{$T_{1} (\unit{\micro\second})$}
        &  \multirow{1}{*}{$T_{2}^\mathrm{SE} (\unit{\micro\second})$}
        & \multirow{1}{*}{$e_\mathrm{sq} (\%)$}
        & \multirow{1}{*}{$e_{0\to1} (\%)$}
        & \multirow{1}{*}{$e_{1\to0} (\%)$}
        \\
        \hline
         $Q_1$ & 38.9 & 15.4 & 0.106 & 0.62&0.70\\
         $Q_2$ & 43.5 & 13.8 & 0.060 & 0.43&0.54\\
         $Q_3$ & 39.7 & 15.3 & 0.075 & 0.41&0.64\\
         $Q_4$ & 39.0 & 13.3 & 0.107 & 0.50&0.90\\
         $Q_5$ & 33.4 & 14.5 & 0.108 & 0.98&0.98\\
         $Q_6$ & 31.0 & 16.2 & 0.119 & 0.34&1.79\\
         $Q_7$ & 37.1 & 18.2 & 0.099 & 0.36&0.56\\
         $Q_8$ & 36.7 & 16.0 & 0.105 & 0.58&1.79\\
         $Q_9$ & 34.0 & 14.2 & 0.100 & 0.33&0.58\\
         $Q_{10}$ &36.2 & 15.8 & 0.090 & 1.94&0.60\\
         $Q_{11}$ &36.6 & 13.7 & 0.123 & 0.75&3.36\\
        \hline
        Average & 36.9 & 15.1 & 0.098 & 0.66&1.13\\
        \hline
        \hline
    \end{tabular}
    \label{tab:T1T2}
\end{table*}

\section{Creating random states}

\begin{figure}
\includegraphics[width=\linewidth]{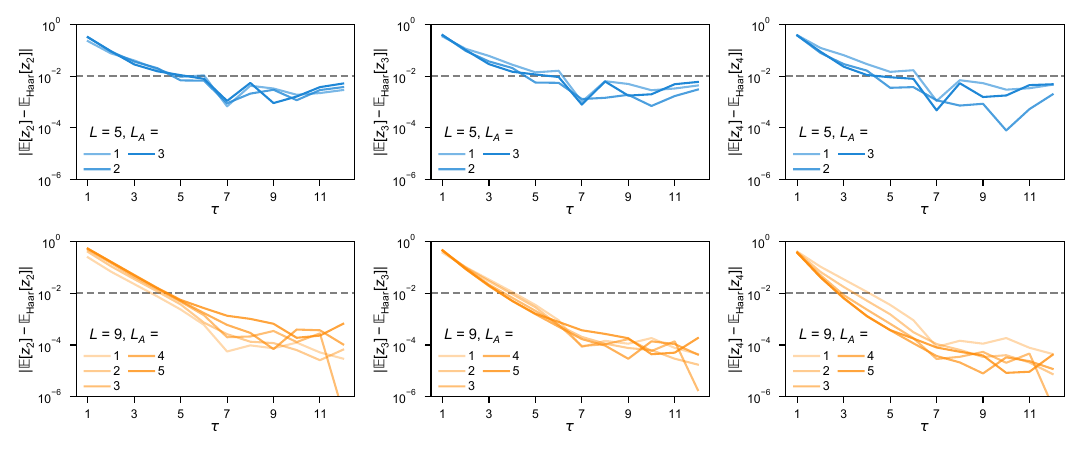}
\caption{
{\bf Efficient preparation of approximate state $k$-designs.} 
Numerical results for the absolute differences between the average reduced density matrix moments in a random state ensemble generated using Eq.~\eqref{eq:designStates_SM} and in the Haar random ensemble. For both cases, we sample $100$ random states to calculate the average moments. The results are shown for total system sizes $L=5$ (top row) and $9$ (bottom row), and different subsystem sizes $L_A$. The difference stays below $10^{-2}$ after the Floquet cycle $\tau \sim 5$ making these approximate state $k$-designs experimentally accessible using low depth Floquet circuits. } 
\label{fig:approxHaar}
\end{figure}

To create random states, the following Floquet evolution is used, 
\begin{equation}
    \ket{\psi}=\left[e^{-i H^{\left(y\right)}T/3} e^{-i H^{\left(z\right)} T /3} e^{-i H^{\left(x\right)} T /3}\right]^\tau \ket{\mathbf{0}},
    \label{eq:designStates_SM}
\end{equation}
with $T$ denoting the Floquet period, taken to be $T=3/J \sim 90~\unit{\nano\second}$ in the experiment, and $\ket{\boldsymbol{0}}=\ket{0}^{\otimes L}$. We prepare up to $30$ independent random states for different system sizes. The constituent Hamiltonians written in the Pauli basis are
\begin{equation}
H^{\left(x, y, z\right)}  = J\sum_{l=1}^{L-1} \left(\sigma^{+}_{l}\sigma^{-}_{l+1} + \sigma^{-}_{l}\sigma^{+}_{l+1}\right) + \sum_{l=1}^{L} h_{l}^{\left(x, y, z\right)} \sigma_{l}^{\left(x, y, z\right)}.
\end{equation}
The local potentials $h_l^{(x, y, z)}$ are drawn uniformly at random from $[-J, J]$. We investigate the optimal number of Floquet cycles, denoted by $\tau$, to reach a good approximation to a Haar random ensemble. To compare the ensemble of states created using this circuit with the Haar random ensemble, we use the notion of approximate state $k$-designs, see for example~\cite{cui2025unitary}. 
A random state ensemble is an $\epsilon$-approximate $k$-design if, for any observable acting on $k$ copies of the state, the average expectation value with respect to the ensemble agrees with that obtained from the Haar ensemble on $\mathbb C^D$, up to an error of at most $\epsilon$. Formally it is written as the trace norm distance between averages of $k$ copy states from the random ensemble and from the Haar ensemble i.e., $||\mathbb{E}(\ket{\psi}\bra{\psi})^{\otimes k}- \mathbb{E}_{\textrm{Haar}}(\ket{\phi}\bra{\phi})^{\otimes k}||_1 \le \epsilon$.  The equivalent definition in terms of traces of moments is, 
\begin{equation}
\left|\mathbb E\left[z_k\right] -\mathbb{E}_{\text{Haar}}\left[z_k\right]\right| \leq \epsilon ~,
\end{equation}
where $z_k=\text{Tr}[\rho_A^k]$ is the $k$-th moment. We numerically compute this difference, as shown in Fig.~\ref{fig:approxHaar}, for the system sizes studied here, and find $\epsilon$ to be $10^{-2}$ for sufficiently large $\tau$ and $k=2$, $3$ and $4$. Notice that, already for $\tau \geq 5$, we have an approximate state $2,3,4$-design. 

Another certification of closeness to Haar-random states comes from fidelity estimation. For pure states $\ket{\phi_s}, \ket{\phi_{s'}}$ sampled from the Haar random ensemble, the average state fidelities are
\begin{equation}
\mathbb{E}_{\text{Haar}}\left[\left|\langle{\phi_{s}}\right|{\phi_{s'}}\rangle|^2\right]=\frac{1}{D}~,
\end{equation}
where $D=2^L$ is the total many-body Hilbert space dimension. In Fig.~1{\bf c} of the the main text, we numerically check that the average fidelity for the states of the form ~\eqref{eq:designStates_SM} converge to $1/D$ with increasing $\tau$. We remark that the states prepared in the experiment are not pure due to decoherence during the Floquet evolution. Thus we use the fidelity between two mixed states $\rho_{\alpha}$ and $\rho_{\beta}$ defined as
\begin{equation}\label{eq:fidelity}
\mathcal{F}(\rho_{\alpha},\rho_{\beta}) = \frac{\mathrm{Tr}\left[\rho_{\alpha} \rho_{\beta}\right]}{\max \left\{\mathrm{Tr}\left[\rho_{\alpha}^2\right], \mathrm{Tr}\left[\rho_{\beta}^2\right]\right\}}~.
\end{equation}
The measurements of the fidelities are done through classical shadows, as presented in Section~\ref{sec:bases}. From these results, it is clear that the prepared ensemble forms an approximate state $k$-design efficiently for $\tau\geq 5$. In the main text and in this SM, we have chosen $\tau=7$ for the preparation of the random ensemble. 

\section{Classical shadows, estimators and error analysis}

In this section, we present the detailed parameter settings for the classical shadow implementation and the post-processing method used to extract various physical observables studied in this work. 

\subsection{Settings of the classical shadow protocol}

The measurement settings are chosen to balance estimation accuracy against the experimental cost of collecting bitstrings and computing the relevant quantities. The specific configurations are summarized in Table~\ref{tab:measurement_settings}, which reports the number of random measurement bases ($M$), the number of measurement shots ($K$), and the number of random states ($R$).

\begin{table*}[!ht]
    \setlength{\tabcolsep}{12pt}
    \renewcommand{\arraystretch}{1.5}
    \centering
    \caption{
    {\bf Measurement settings.} For each figure in the main text, we list the number of random measurement bases ($M$), the number of measurement shots ($K$), and the number of random states ($R$). }
    \begin{tabular}{c|c|c|c|c}
        \hline \hline
        \multirow{1}{*}{Figure}
        & \multirow{1}{*}{$L$}
        &  \multirow{1}{*}{$M$}
        &  \multirow{1}{*}{$K$}
        & \multirow{1}{*}{$R$}
        \\
        \hline
         Fig. 1 & 5 & 1000 & 5000 & 20\\
          &7 & 1000 & 5000 & 20\\
          \hline
         \multirow{4}{*}{Fig. 2}& 5& 1000 & 10000 & 30\\
          & 7& 1000 & 10000 & 30\\
         & 9& 1000 & 10000 & 20\\
         & 11& 1000 & 10000 & 20\\
         \hline
         Fig. 3& 9& 1000 & 10000 & 20\\
         \hline
         Fig. 4& 7& 1000 & 5000 & 20\\
        \hline
        \hline
    \end{tabular}
     \label{tab:measurement_settings}
\end{table*}

\subsection{Number of measurement bases}\label{sec:bases}

\begin{figure}[!t]
    \centering
    \includegraphics[width=\linewidth]{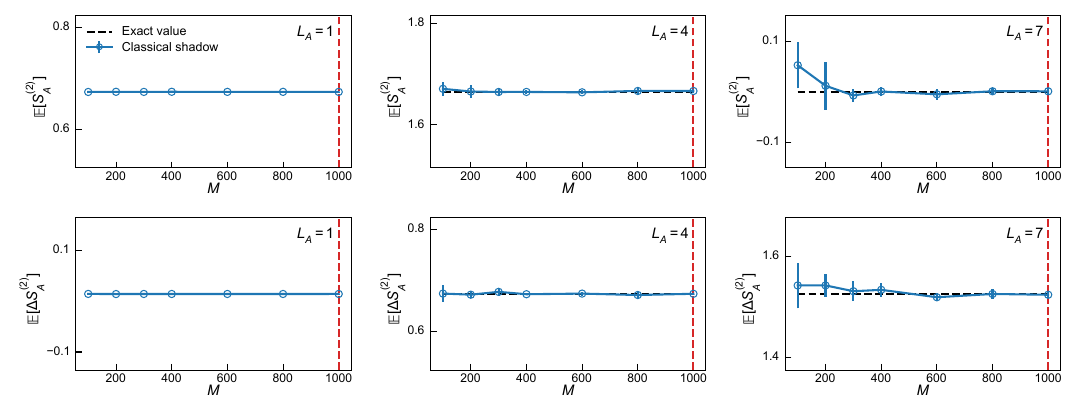}
    \caption{{\bf Statistical uncertainty and estimation bias of classical shadows.} 
    In these noiseless numerical simulations, we estimate the average R\'enyi-2 entropy (top panel) and the entanglement asymmetry (bottom panel) of $30$ random states using batch shadows, varying the number $M$ of random measurement bases. Each simulation is repeated 6 times to compute the mean values (data points) and the corresponding standard deviations (error bars). 
    The horizontal dashed lines are the exact average entanglement entropy and asymmetry of the random states generated in the simulation. 
    The red vertical dashed line indicates the value of $M$ used in our experiment. 
    }
    \label{figS:RE_VS_M}
\end{figure}

For the classical shadow protocol, it is crucial to determine an appropriate number of random measurement bases ($M$), since an insufficient number of measurement bases can result in significant statistical uncertainty and estimation bias. To address this, we perform noiseless numerical simulations on a system of size $L=7$, We first generate an ensemble of $30$ random states via Floquet dynamics. For each state, we randomly sample $M$ measurement bases with replacement (after each draw, the selected basis is returned to the pool before the next draw). We then compute the exact classical shadow corresponding to the $m$-th measurement basis with the formula
\begin{equation}
    \rho^{(m)} = \sum_{s}P(s)\bigotimes_{l=1}^{L} \left(3(U^{(m)}_{l})^{\dagger}|s_{l}\rangle\langle s_{l}|U_{l}^{(m)} - \mathbb{I} \right),
    \label{eq:classical_shadow}
\end{equation}
where $P(s)$ denotes the probability of measuring the bitstring $s$. Using these classical shadows, we construct $B=20$ batch shadows to estimate the average R\'enyi-2 entanglement entropy and the entanglement asymmetry. This procedure is repeated over $6$ independent realizations to quantify the standard deviation, which serves as a measure of statistical uncertainty due to finiteness of $M$. As illustrated in Fig.~\ref{figS:RE_VS_M}, choosing $M=1000$ random measurement bases is sufficient to obtain accurate and converged results.

In the case of the fidelity~\eqref{eq:fidelity}, to estimate the state overlap $\mathrm{Tr}\left[\rho_{\alpha} \rho_{\beta}\right]$ in the experiment, we employ $M$ randomly selected measurement bases to construct the classical shadows, which are shared across all $20$ random states. The overlap is accurately estimated by 
\begin{equation}
\mathrm{Tr}\left[\rho_{\alpha} \rho_{\beta}\right] = \frac{1}{M\left(M-1\right)} \sum_{m \neq m'} \mathrm{Tr} \left[\rho_{\alpha}^{(m)} \rho_{\beta}^{(m')}\right],
\end{equation}
where $\rho_{\alpha}^{(m)}$ denotes the $m$-th classical shadow of $\rho_{\alpha}$.
Meanwhile, the purities $\mathrm{Tr}\left[\rho_{\alpha}^2\right]$ and $\mathrm{Tr}\left[\rho_{\beta}^2\right]$ can be obtained using the same shadows, as explained in the Methods section of the main text. The results of the fidelity estimation are shown in Fig.~1{\bf c} of the main text.

\subsection{Readout error correction}

Measurements of superconducting qubits are vulnerable to various types of errors, which manifest as readout errors. There is a probability of measuring the state $\ket{k}$ when the qubit is actually in the state $\ket{j}$. This affects the probabilities of the measured bitstrings used in the construction of the classical shadows, as reported in Eq.~\eqref{eq:classical_shadow}.
To this end, we apply a readout error correction scheme to the classical shadow formalism. We first perform a calibration experiment to obtain the response matrix~\cite{Bravyi2021Phys.Rev.A}. By inverting this matrix, we reconstruct the corrected bitstring probabilities from the measured probabilities~\cite{PhysRevLett.118.210504}. The corrected classical shadow is then calculated by using the corrected bitstring probabilities in Eq.~\eqref{eq:classical_shadow}.

\begin{figure}[t]
    \centering
    \includegraphics[width=\linewidth]{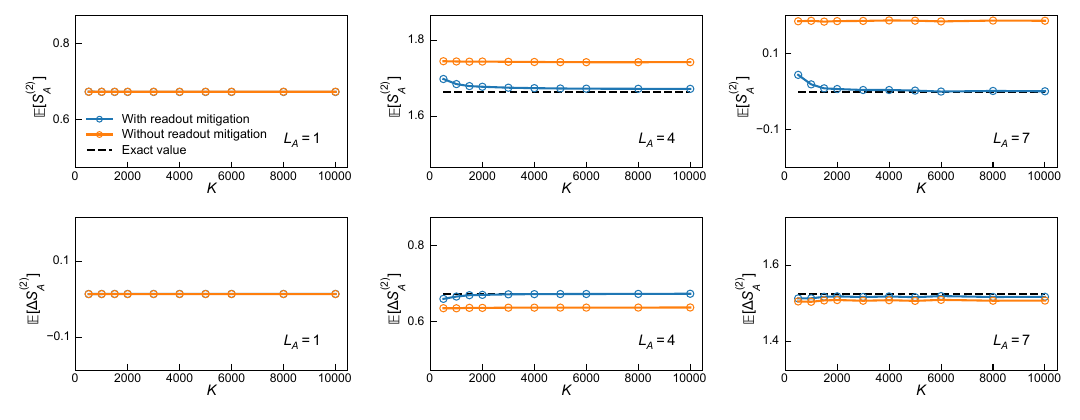}
    \caption{
    \textbf{Numerical verification of the efficacy of the readout error correction.} 
    In this simulation, we estimate the average R\'enyi-2 entropy (top panel) and entanglement asymmetry (bottom panel) of 30 random states in a $L=7$ qubit system subject to readout error rates of $e_{0\to1}=0.007$ and $e_{1\to0}=0.011$. We take subsystem sizes $L_A = 1, 4, 7$ and vary the number $K$ of shots per basis for $M=1000$ random bases. The corrected estimates (blue circles) are more accurate than the uncorrected estimates (orange circles). As $K$ increases, the corrected estimates converge toward the exact average entanglement entropy and asymmetry of the random states considered  (black dashed lines). 
    }
    \label{figS:RE_vs_K}
\end{figure}

We validate the efficacy of this readout error correction strategy via numerical simulations, as shown in Fig.~\ref{figS:RE_vs_K}. The simulation estimates the average R\'enyi-2 entropy and entanglement asymmetry for a $7$-qubit system subject to readout error rates of $e_{0\to1}=0.007$ and $e_{1\to0}=0.011$. Here, we fix the number of random states at $R=30$, the number of measurement bases at $M=1000$, and the number of batch shadows at $B=20$. The results demonstrate that the corrected estimates exhibit significantly reduced bias and converge toward the exact values as the number of shots per basis ($K$) increases. Based on these results, we select $K=5000$ and $K=10000$ for the experiments discussed in the main text.

\subsection{Decoherence and error mitigation}

Decoherence during the Floquet evolution~\eqref{eq:designStates_SM} causes the experimentally generated random states to be non-pure, leading to a notable deviation between the experimental estimates and the theoretical predictions.
In fact, by independently characterizing the decoherent times (see Table~\ref{tab:T1T2}) and simulating the open-system dynamics via the Gorini-Kossakowski-Sudarshan-Lindblad master equation, we obtain numerical results that are in close agreement with the experimental results. 

The Floquet operator $V$ creates an ergodic dynamics with CUE-like eigenspectrum~\cite{Dong2025Phys.Rev.Lett.}. For the discussion in this section, we assume that the unitary $V$ exhibits perfect Haar-random behavior, despite the finite size of the system. Under the evolution in the presence of decoherence, the state changes from $\rho$ to $\Lambda(V\rho V^\dagger)$, where $\Lambda$ denotes a quantum channel which includes the unitary operation $V$ and possible interactions with the environment.  We assume the errors to be gate- and time-independent, and we model the resulting noise channel as a completely positive trace-preserving (CPTP) map acting on the state $\rho$. When considering averaged quantities, such channels are known to effectively behave like a depolarization channel when the unitary $V$ is drawn from a Haar random ensemble on the unitary group $U(D)$~\cite{Mele2024introductiontohaar}. Thus, to mitigate the decoherence effects, we take the resulting noisy density matrix $\rho^{(\mathrm{deco})}$ to be related to the ideal noiseless density matrix $\rho$ by
\begin{equation}
    \rho^{(\mathrm{deco})} = \left(1-\epsilon\right)\rho + \epsilon\,\frac{\mathbb{I}}{D},
    \label{eq:depolarizing_channel}
\end{equation} 
where $\epsilon$ denotes the depolarizing strength and $\mathbb{I}$ is the $D\times D$ identity matrix. 
Based on this model, the noisy $n$-th order moment of the density matrix, defined as $z_{n}^{\mathrm{(deco)}} = \mathrm{Tr}\left[\left(\rho^{\mathrm{(deco)}}\right)^n\right]$, can be expressed as a linear combination of the noiseless moments $\{z_1, \dots, z_n\}$. Applying the binomial expansion to Eq.~\eqref{eq:depolarizing_channel}, we obtain
\begin{equation}\label{eq:z_n_depol}
    z_{n}^{\mathrm{(deco)}} = \sum_{k=1}^{n} \binom{n}{k} \left(1 - \epsilon\right)^k \left(\frac{\epsilon}{D}\right)^{n-k} z_k + \dfrac{\epsilon^n}{D^{n-1}}.
\end{equation}
We note that $\rho^{({\rm deco})}$, like any density matrix, is normalized, so that its first moment $z_{1}^{\mathrm{(deco)}} = 1$, independent of $\epsilon$. For the specific case $n=2$, the relation~\eqref{eq:z_n_depol} between $z_n$ and $z_n^{({\rm deco})}$ can be rewritten in the matrix form as
\begin{equation}
    \begin{bmatrix}
        z^{\mathrm{(deco)}}_1 \\ z^{\mathrm{(deco)}}_2
    \end{bmatrix}
    = 
    \mathbf{M}_2 
    \begin{bmatrix}
        z_1 \\ z_2
    \end{bmatrix},
\end{equation}
where the decoherence response matrix $\mathbf{M}_2$ is given by
\begin{equation}
    \mathbf{M}_2 = 
    \begin{bmatrix}
        1 & 0 \\
        \dfrac{\epsilon\left(2 - \epsilon\right)}{D} & \left(1 - \epsilon\right)^2
    \end{bmatrix}.
    \label{eq:decoherence_response_matrix}
\end{equation}

The error mitigation protocol is implemented as follows. First, the experimental set of generated random states is partitioned into two subsets, each containing $R/2$ random states. The first subset is used to estimate the depolarizing strength $\epsilon$. The other subset is employed to estimate the error-corrected average entropy, entanglement asymmetry and partial-transpose moments. Assuming that the ideal random states are pure, i.e., $z_2=1$, $\epsilon$ can be extracted from the average $\mathbb{E}\left[p^{\mathrm{(exp)}}_2\right]$ of the second moments of the first $R/2$ experimental random states via~\cite{Joshi2022Phys.Rev.X}
\begin{equation}\label{eq:dep_strength}
    \epsilon = 1 - \sqrt{\frac{D \, \mathbb{E}\left[z^{\mathrm{(exp)}}_2\right] - 1}{D-1}}.
\end{equation}
Next, we construct the decoherence response matrix $\mathbf{M}_2$ utilizing Eq.~\eqref{eq:decoherence_response_matrix}. Finally, the error-mitigated moments $z^{\mathrm{(em)}}_2$ are obtained from the experimental ones $z^{\mathrm{(exp)}}_2$ by inverting the response matrix:
\begin{equation}\label{eq:inv_resp_mat}
    \begin{bmatrix}
        z^{\mathrm{(em)}}_1 \\ z^{\mathrm{(em)}}_2
    \end{bmatrix}
    = \mathbf{M}_2^{-1} 
    \begin{bmatrix}
        z^{\mathrm{(exp)}}_1 \\ z^{\mathrm{(exp)}}_2
    \end{bmatrix}.
\end{equation}
Due to the linearity of the partial trace, the same relation applies to the moments of the reduced density matrix.
For the entanglement asymmetry, the error-mitigated second moments of the symmetrized density matrix $\rho_{A,Q}$  are also determined by applying Eq.~\eqref{eq:inv_resp_mat}, since the projection onto the symmetry sectors is a linear operation.
This protocol can be readily extended to higher-order moments to compute the error-mitigated moment ratio $\tilde{r}_2$.
In all cases, we use the same value of $\epsilon$ extracted from Eq.~\eqref{eq:dep_strength}.

\subsection{Estimation of error bars}

\begin{figure}[!t]
    \centering
    \includegraphics[width=\linewidth]{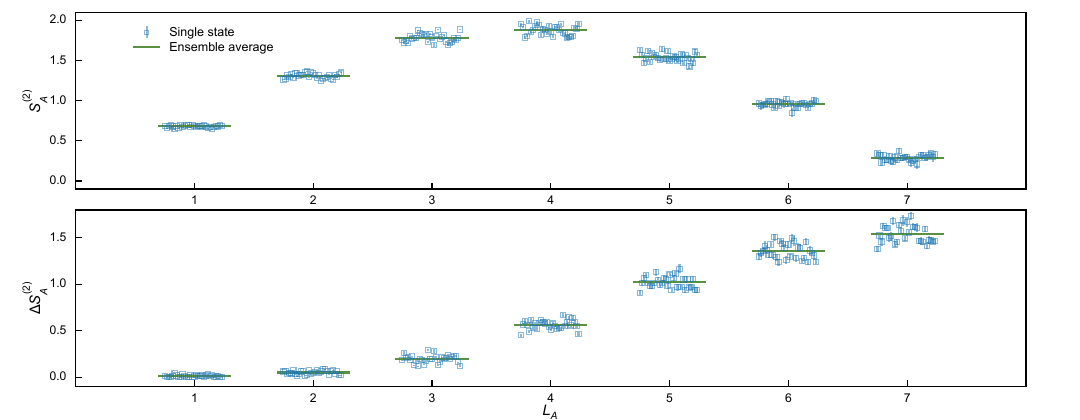}
    \caption{{\bf Statistical uncertainties of the classical shadow estimates.}
    Symbols are the subsystem R\'enyi-2 entropies (top panel) and entanglement asymmetries (bottom panel) for a single random state in a system of $L=7$ qubits. For each data point, the error bars are the statistical uncertainties estimated using the jackknife resampling method over random measurements. The solid green lines represent the average value. All data points shown here are experimental results without any error-mitigation.
    }
    \label{figS:JK_VS_SEM}
\end{figure}

In Fig.~\ref{figS:JK_VS_SEM}, we present the statistical uncertainties of the classical shadow estimates for each individual random state generated in a $7$-qubit system, as obtained via the jackknife resampling method~\cite{Satzinger2021Science}. These uncertainties are small, owing to the selection of a significant number of random measurement bases. Therefore, the Fig.~2 in the main text employ error bars representing the standard error of the mean (SEM) over finite number of random states to indicate the accuracy of the ensemble average, and the dominant error.
In contrast, the error bar in Fig.~3 are estimated using the jackknife resampling technique, implemented as follows. For an observable $O$, we compute the $r$-th jackknife sample $\bar{O}_r$ by excluding the measurement result corresponding to the $r$-th random state. This process produces $R$ jackknife samples $\{\bar{O}_1, \cdots, \bar{O}_R\}$. The final estimate of the observable is then given by
\begin{equation} 
\bar{O} = \frac{1}{R}\sum_{r=1}^R \bar{O}_r\,,
\end{equation}
and the error bar is computed as
\begin{equation} 
\sigma = \sqrt{\frac{R-1}{R}\sum_{r=1}^R \left(\bar{O}_r-\bar{O}\right)}\,. 
\end{equation}

\section{Analytic observable averages in the Haar-random state ensemble}

In the main text, we compare the average R\'enyi-2 entanglement entropy, the entanglement asymmetry, and the partial-transpose moments of the random states generated experimentally using the Floquet circuit~\eqref{eq:designStates_SM} with those expected for Haar-random states. For this ensemble, there exist analytic formulas for these observables, which can be derived using random-matrix techniques such as Weingarten calculus~\cite{Collins2016}. 
Here, we collect the explicit formulas plotted in the figures of the main text and this Supplementary Material.

In the case of the R\'enyi-2 entanglement entropy and entanglement asymmetry of a subsystem $A$, we used the expressions obtained for them in Ref.~\cite{Ares2024Phys.Rev.D} at finite $L$. The only approximation applied in their derivation is 
\begin{equation}
\mathbb{E}_{{\rm Haar}}\left[\log\left({\rm Tr}\left(\rho_A^2\right)\right)\right]\approx \log\left(\mathbb{E}_{{\rm Haar}}\left[{\rm Tr}\left(\rho_A^2\right)\right]\right);
\end{equation}
that is, introducing the average inside the logarithm. The same approximation is applied to the moments of the symmetrized reduced density matrix $\rho_{A, Q}$, which enters the definition of the entanglement asymmetry. In the Haar-random ensemble, this approximation becomes exact in the thermodynamic limit $L\to\infty$; at finite $L$, it gives rise to exponentially small corrections, as shown in e.g. Ref.~\cite{Russotto2025NonAbelian}.
Up to these negligible corrections, we have
\begin{equation}
\mathbb{E}_{\rm Haar}[S_A^{(2)}]=  -\log\left(\frac{2^{L - L_A} + 2^{L_A}}{2^L + 1}\right)
\end{equation}
for the R\'enyi-2 entanglement entropy and 
\begin{equation}
\mathbb{E}_{\rm Haar}[\Delta S_A^{(2)}]= -\log\left[\frac{1}{2^{2L_A-L} +1}\left(1+ 2^{-L_A} \frac{(2L_A)!}{(L_A!)^2}\right)\right]
\end{equation}
for the entanglement asymmetry.

For the moments $p_k={\rm Tr}[(\rho^\Gamma)^k]$ of the partially transposed reduced density matrix $\rho^{\Gamma} = \left(\mathbb{I}_A \otimes T_B\right) \text{Tr}_C[\rho]$, we use the following exact combinatorial expression obtained in Refs.~\cite{Banica2013,PRXQuantum.2.030347},
\begin{equation}
\mathbb{E}_{\rm Haar}[p_k]=\frac{1}{(L_AL_BL_C)^k}\sum_{\tau\in S_k} L_C^{c(\tau)}L_A^{c(\sigma_+\circ\tau)}L_B^{c(\sigma_-\circ\tau)},
\end{equation}
where $S_k$ is the permutation group of degree $k$ and $\sigma_\pm$ are the cyclic and anti-cyclic permutations of $k$ elements, i.e. $\sigma\pm (j)=j\pm 1$ $({\rm mod}\,k)$. For a given permutation $\tau\in S_k$, $c(\tau)$ is the number of cycles in the permutation, including the trivial ones.

\section{Extended experimental results}

\subsection{Phase diagram of moment ratio for $L=11$}

\begin{figure}[!t]
    \centering
    \includegraphics[width=\linewidth]{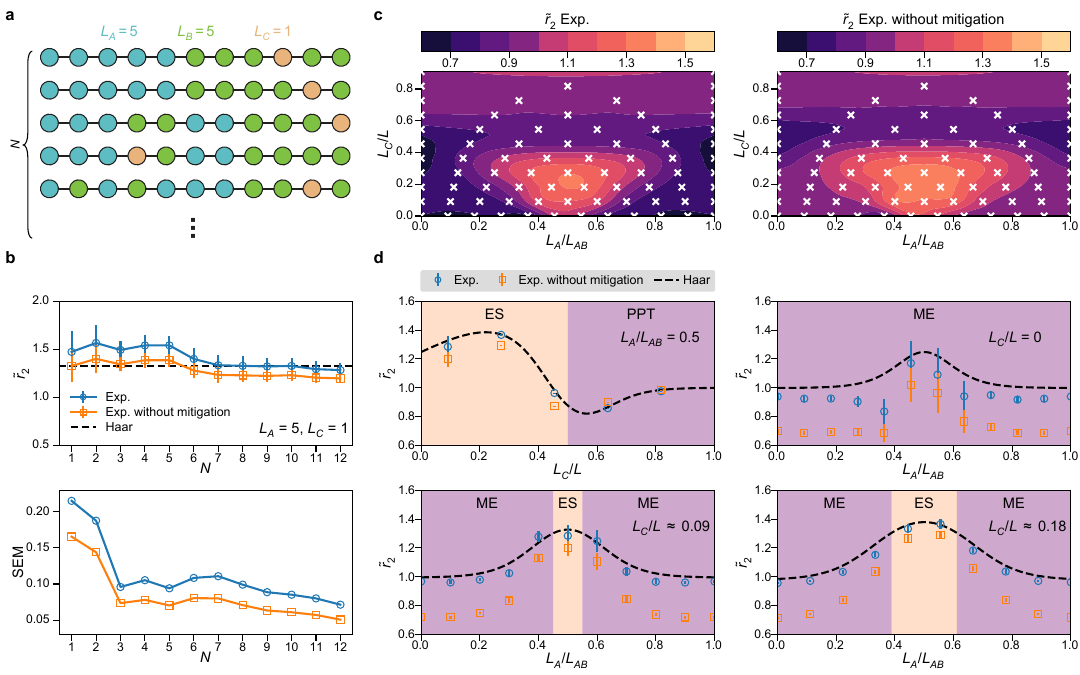}
    \caption{{\bf Entanglement phase diagram of random states characterized by the moment ratio $\tilde{r}_2$ for $L=11$.} 
    {\bf a}, Different partitionings of the qubit chain into three subsystems for a fixed configuration of subsystem sizes. In this example, we fixed $L_A=5$, $L_B=5$, and $L_C=1$. 
    {\bf b}, Estimates and SEM of the moment ratio $\tilde{r}_2$ averaged over different number $N$ of partitions. As $N$ increases, the averaged values converge toward the theoretical prediction, and the SEM, estimated using the jackknife resampling method, are reduced. 
    {\bf c}, Raw (left panel) and error-mitigated (right panel) entanglement phase diagrams. In both cases, white crosses are the specific coordinates of the experimental data points used to draw the colored contour plot by cubic interpolation. 
    {\bf d}, Experimental results and theoretical predictions with fixed partition ratios of $L_A/L_{AB}=0.5$, $L_C/L=0$, $L_C/L\approx0.09$, and $L_C/L\approx0.18$. Circles and squares represent error-mitigated and raw data, respectively, while dashed lines are the theoretical prediction for Haar random states.
    }
    \label{figS:L=11_r2}
\end{figure}

Due to the limited number of random measurements ($M = 1000$), the estimation of $\tilde{r}_2$ for $L = 11$ becomes less reliable, particularly at $L_C/L = 0$. The statistics can be improved by averaging over different partitions for each individual state in the evaluation of the $k$th-order moment of the partially transposed state, $p_k$. As shown in Fig.~\ref{figS:L=11_r2}\textbf{a}, for a fixed subsystem configuration $\{L_A, L_B, L_C\}$, we consider up to $N$ distinct partitions. Note that, for certain subsystem configurations, the number of available partitions may be less than $N$. In such cases, we take the average over all available partitions.
In Fig.~\ref{figS:L=11_r2}\textbf{b}, we demonstrate that, with increasing $N$, the $\tilde{r}_2$ converges toward the theoretical prediction, accompanied by a reduction in the SEM estimated via the jackknife resampling method. After averaging over up to $N = 12$ different partitions for each individual state, we present the entanglement phase diagram characterized by the moment ratio $\tilde{r}_2$ for system size $L = 11$ in Fig.~\ref{figS:L=11_r2}\textbf{c}. Vertical and horizontal slices of the entanglement phase diagram are shown in Fig.~\ref{figS:L=11_r2}\textbf{d}, providing a more detailed view of the transitions between the  maximally entangled (ME), entanglement saturation (ES), and the positive partial transpose (PPT) states in the $L = 11$ qubit system. 

\subsection{Phase diagram of negativity}

\begin{figure}[!t]
    \centering
    \includegraphics[width=\linewidth]{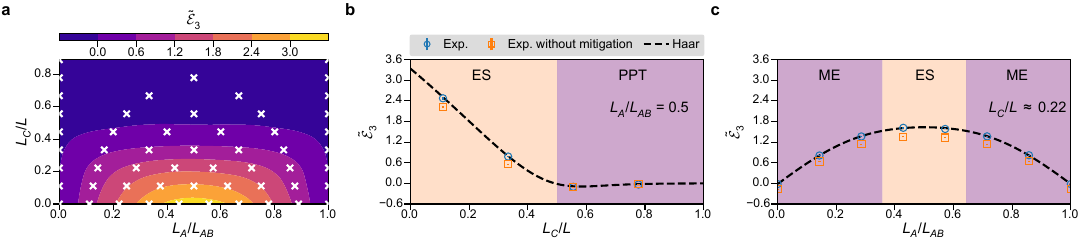}
    \caption{\textbf{Entanglement phase diagram of random states characterized by the $p_3$-negativity $\tilde{\mathcal{E}}_3$ for $L=9$.} 
    \textbf{a}, Experimental phase diagram of the $p_3$-negativity. White crosses denote the specific coordinates of the error-mitigated experimental data points used to draw the colored contour plot by cubic interpolation. 
    \textbf{b}--\textbf{c}, Experimental results and theoretical prediction for Haar-random states (dashed curves) with fixed partition ratios at $L_A/L_{AB}=0.5$ (\textbf{b}) and $L_C/L\approx0.22$ (\textbf{c}). Circles and squares represent error-mitigated and raw data, respectively.
    }
    \label{figS:p3_negativity}
\end{figure}

To distinguish the ME phase from the PPT phases, in $\tilde{r}_2$ takes the same value, we study the $p_3$-negativity of random states~\cite{r2_phases}, defined as
\begin{equation}
    \tilde{\mathcal{E}}_3 = \frac{1}{2} \log_2\left(\frac{\mathbb{E}[p_2]^2}{\mathbb{E}[p_3]}\right),
\end{equation}
where $p_n$ is $n$th-order moment of the partially transposed state (see the main text).
In Fig.~\ref{figS:p3_negativity}\textbf{a}, we plot $\tilde{\mathcal{E}}_3$ for a system of size $L=9$, using the same dataset as in Fig.~4 of the main text. We observe distinct behaviors across the different phases. In the PPT phase, $\tilde{\mathcal{E}}_3$ remains $0$, independent of the subsystem sizes $L_A$ and $L_C$. In the entanglement saturation (ES) phase, $\tilde{\mathcal{E}}_3$ decreases monotonically as the environment size $L_C$ increases (Fig.~\ref{figS:p3_negativity}\textbf{b}). In the ME phase, $\tilde{\mathcal{E}}_3$ increases with the subsystem size $L_A$ while $L_A < L_B$, and subsequently decreases symmetrically when $L_A > L_B$ (Fig.~\ref{figS:p3_negativity}\textbf{c}). 
In the absence of environment ($L_C = 0$), the moments of the partial transpose coincide with the moments of the  density matrix $\rho_A$ such that $p_2=\left({\rm Tr}\left[\rho_A\right]\right)^2=1$ and $p_3={\rm Tr}\left[\rho_A^3\right]$, see Ref.~\cite{Calabrese12Negativity}. 
Therefore, the $p_3$-negativity reduces to $\tilde{\mathcal{E}_3}=-\left(\log_2\mathbb{E}\left[{\rm Tr}\left[\rho_A^3\right]\right]\right)/2$
and, consequently, it exhibits a Page-curve behavior as a function of $L_A$.

\bibliography{references}